\documentstyle[proceedings,graphicx]{crckapb}
\begin{opening}
  \title{THE INFRARED [WC] STARS \\
}
\author{Albert A. Zijlstra}
\institute{The University of Manchester\\
  Manchester M13 9PL, UK }
\end{opening}
\begin{document}

\def\msol{{M_\odot}}

\noindent (First published in Astrophysics \&\ Space Science, 275, 90
(2001). Some references to more recent works have been added as
footnotes to the original text)
\vskip 12pt

\begin{abstract}
A number of late [WC] stars have unique infrared properties, not found
among the non-[WC] planetary nebulae, and together define a class of IR-[WC]
stars. They have unusual IRAS colours, resembling stars in the
earliest post-AGB evolution and possibly related to PAH formation.
Most or all show a double chemistry, with both a neutral (molecular)
oxygen-rich and an inner carbon-rich region. Their dense nebulae indicate
recent evolution from the AGB, suggesting a fatal-thermal-pulse (FTP)
scenario. Although both the colours and the stellar characteristics
predict fast evolution, it is shown that this phase must last for
$10^4\,\rm yr$. The morphologies of the nebulae are discussed.  For
one object in Sgr, the progenitor mass ($1.3\, \rm M_\odot$) is known.
The stellar temperatures of the IR-[WC] stars appear much higher in
low metallicity systems (LMC, Sgr).  This may be indicative of an
extended 'pseudo' photosphere.  It is proposed that re-accretion of
ejected gas may slow down the post-AGB evolution and so extend the life
time of the IR-[WC] stars.

\end{abstract} 
\section{Introduction}

The nebulae surrounding the hydrogen-poor emission-line [WC] stars are
thought not to differ from those surrounding non-[WC] stars, except for
a higher expansion velocity (Gorny \&\ Stasinska 1995) and higher
internal turbulence (Gesicki et al. 1998), both of which can be
understood as the effect of the hydrogen-poor wind on the nebula
(Mellema, these proceedings).  The lack of chemical peculiarities in the
nebulae shows that the  ejected occured before the star
became hydrogen poor. The normal range of ionized masses indicates
that this occured on the AGB, in a normal AGB superwind phase.

The hydrogen-poor nature of the [WC] star is probably related to a
thermal pulse. This can have occured during three possible phases:
(1) on the white dwarf cooling track (a very late thermal pulse or
VLTP); (2) during the hydrogen-burning post-AGB evolution (a late
thermal pulse or LTP); (3) at the very end of the AGB, with the pulse
terminating the AGB evolution (a fatal thermal pulse or FTP).  The
VLTP model is preferred from stellar modelling (e.g. Herwig et
al. 1999). The nebulae surrounding VLTP [WC] stars should be highly
evolved, since the pulse occurs $\sim 10^4\,\rm yr$ after the AGB.  The
envelope mass of the star prior to VLTP is very small ($10^{-4} \msol$), and so
the PN cannot be resupplied with significant amounts of hydrogen-rich
gas.  Some of the [WC] PNe are so compact that they must have been
ejected on the AGB very recently. For these objects the FTP scenario
may be preferred (Zijlstra et al. 1991).

The compact [WC] PNe are extremely bright infrared emitters. I will
show that their nebulae have unexpected characteristics, and do not
have a counterpart among non-[WC] PN.  \footnote{For a general recent
  review of planetary nebulae, published after these proceedings, see
  Kwitter \&\ Henry 2022. A list of Galactic [WC] stars is in
  Muthumariappan \&\ Parthasarathy 2020}

% Several planetary nebulae around late [WC] stars are now known to be
% surrounded by both an oxygen-rich and a carbon-rich dust shell,
% showing evidence for chemical evolution (Zijlstra et al. 1991, Waters
% et al, 1998, Cohen et al. 1998).  

\section{Infrared flux and colours}

\begin{figure}
\includegraphics[width=125mm,height=125mm]{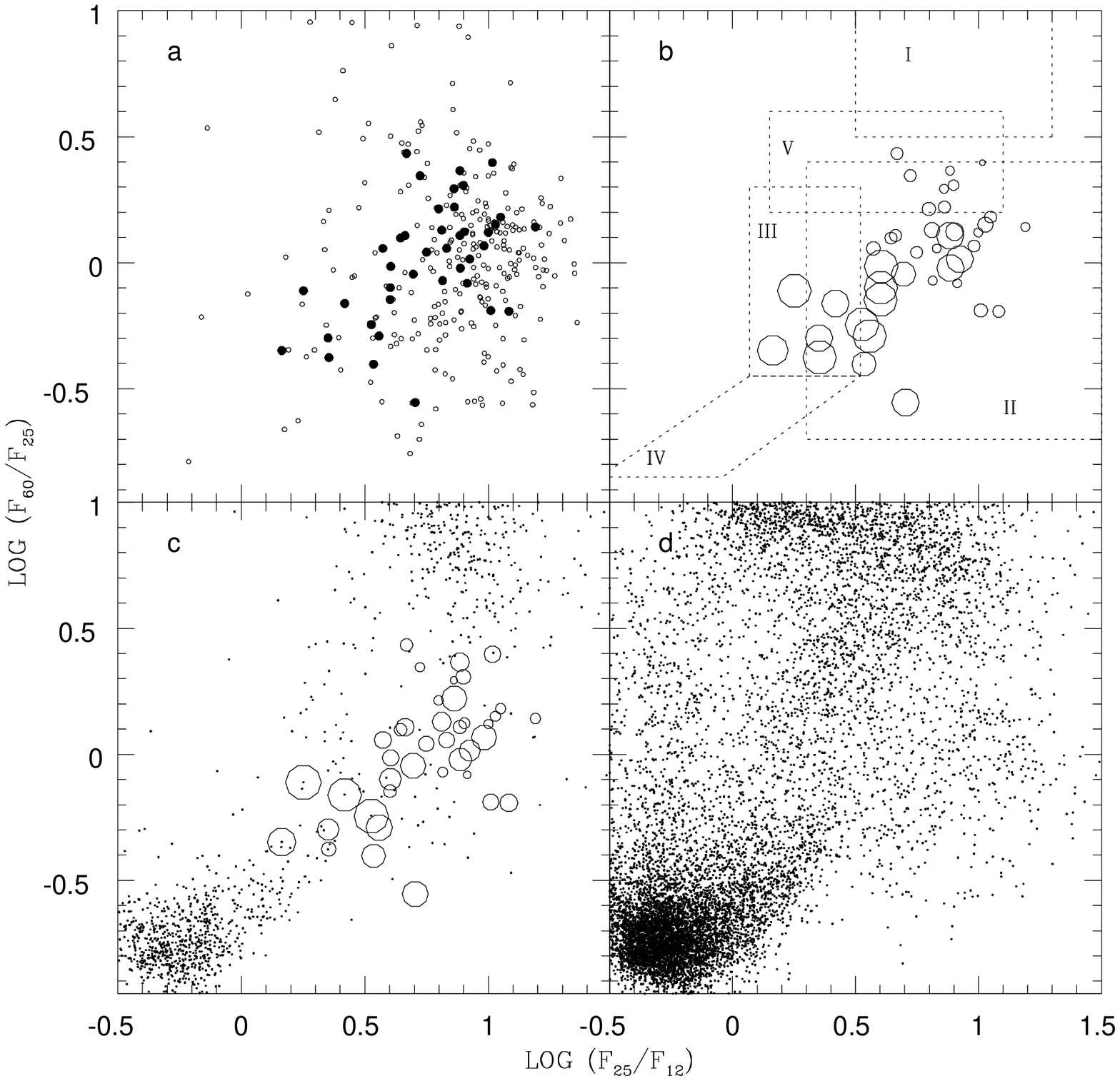}
%\vspace{15cm}  % amount of vertical space needed
\caption{Infrared IRAS colours. The axes are defined in terms of
12$\mu$m, 25$\mu$m and 60$\mu$m IRAS flux densities, without colour
correction.  Panel a: the open circles show all PNe detected in all
three bands; the filled circles are all the [WC] stars. Panel b: the
[WC] stars; the size of the circles is proportional to the subclass
(class 11 are the largest circles). The boxes show the expected
location of HII regions (I), PNe (II), PostAGB stars (III), Miras (IV)
and bipolar outflow sources (V) (from Zijlstra et al. 2001).  Panel c:
the open circles are the [WC] stars; the size of the symbols is now
proportional to the integrated IRAS flux. The dots represent all
IRAS sources with integrated flux comparable  to the IR-brightest [WC]
stars. Panel d: All IRAS sources with good detections in all three
bands.}
\end{figure}

Table 1 lists all known Galactic [WC] stars in the catalog of Gorny
(these proceedings), with IRAS detections (flux quality 2 or 3) at 12,
25 and 60 micron.  IRAS21282+5050 is included: although this object
does not satisfy the [WC] criteria due to the lack of higher
ionization levels (Crowther et al. 1998), it shows characteristics
very similar to belong to the late [WC] stars.  PC14 is not included:
it has very peculiar IRAS colours but the IRAS position is more than 1
arcmin from the PN and confusion is likely.  The total sample contains 40
objects. There is no excess of objects towards the Bulge suggesting all are
Galactic disk objects.

\begin{table}[p]
\begin{center}
\caption{IRAS data for Galactic [WC] stars, in order of Galactic longitude}
\begin{tabular}{lllllll}
\hline
Name & [WC] subtype & $F_{12}$ & $F_{25}$ &  $F_{60}$  \\
     &              &  [Jy]    & [Jy]     &  [Jy]      \\
\hline
SwSt1             &   9   & 17    &  86   &  24    & IR-[WC]? \\ %  1.5-6.7
Cn1-5             &   4   & 1.7   &  7.5  &   9.4  \\ %  2.2-9.4
NGC6369           &   4   & 9.0   & 65.5  & 109    \\ %  2.5+5.8
Hb4               &   3.5 & 1.3   & 10.3  & 20.9   \\ %  3.1+2.9
M1-25             &   6   & 0.8   &  6.4  &  8.5   \\ %  4.9+4.9
M3-15             &   5   & 0.53  &  5.66 &  8.02  \\ %  6.8+4.1
M1-32             &   4.5 & 4.2   & 15.7  & 17.9   \\ % 11.9+4.2
PM1-188           &  11   & 3.8   & 15.3  & 14.8   \\ % 12.2+4.9
M1-61             &   4   &  2.19 &  26.5 &  17.0  \\ %  19.4-5.3
M1-60             &   4   &   0.5 &   5.6 &  8.5   \\ % 19.7-4.5
M1-51             &   5   &   3.6 &  23.3 & 31.4   \\ % 20.9-1.1
M2-43             &   8   & 15.7  & 53.8  & 21.3   & IR-[WC]? \\ % 27.6+4.2
NGC6751           &   4   & 4.0   & 18.4  & 23.6   \\ % 29.2-5.9
He2-429           &   4.5 & 0.7   &  4.4  &  7.2   \\ % 48.7+1.9
PM1-310           &  11   & 2.1   &  8.4  &  6.0   \\ % 60.4+1.5
NGC6905           &   3   & 0.4   &  6.2  &  8.6   \\ % 61.4-9.5
BD+30$\rm ^o$3639 &   9   & 89.4  & 234.5 & 161.7  & IR-[WC] \\ % 64.7+5.0
He2-459           &   9   & 4.5   & 37.8  & 39.1   \\ % 68.3-2.7
NGC7026           &   3   & 2.4   & 18.4  & 42.7   \\ % 89.0+0.3
IRAS21282+5050    &  10   & 51.0  &  74.4 &  33.4   & IR-[WC] \\ %
NGC40             &   8   & 14.5  & 71.9  & 64.8   \\ %120.0+9.8 
IC1747            &   4   & 0.7   &  3.7  &  8.2   \\ %130.2+1.3
NGC1501           &   4   & 1.2   &  5.6  & 15.2   \\ %144.5+6.5
M4-18             &  11   & 4.2   &  9.5  &  4.0    & IR-[WC]? \\ %146.7+7.6
IC2003            &   3   & 0.5   &  4.1  &  3.4   \\ %161.2-14.8
NGC2452           &   3   & 0.5   &  5.0  &  6.6   \\ %243.3-1.0
NGC2867           &   3   & 2.2   & 14.9  & 17.0   \\ %278.1-5.9
PB6               &   3   & 0.92  &  6.0  &  5.1   \\ %278.8+4.9
Pe1-1             &   4.5 & 1.8   & 18.4  & 11.9   \\ %285.4+1.5
IRAS07027$-$7934  &  11   & 22.7  & 82.0  & 42.0    & IR-[WC] \\ %291.3-26.2
NGC5189           &   2   & 1.3   & 13.5  & 33.7   \\ %307.2-3.4
He2-99            &   9   & 1.2   &  9.2  & 11.8   \\ %309.0-4.2
NGC5315           &   4   & 7.5   & 72.0  & 84.0   \\ %309.1-4.3
He2-113           &  11   & 92.4  & 310.5 & 176.6   & IR-[WC] \\ %321.0+3.9
PM1-89            &   4   & 2.3   & 12.9  & 14.2   \\ %324.0+3.5
He2-142           &   9   & 13.8  & 31.0  & 15.6   & IR-[WC]? \\ %327.1-2.2
He3-1333          &  11   & 144.  & 257.  & 199.    & IR-[WC] \\ %332.9-9.9
% PC14              &   5   & 2.9   &  2.0  &  3.4   \\ %336.2-6.9
Pe1-7             &   9   & 6.5   & 50.0  & 47.6   \\ %337.4+1.6
K2-16             &  11   & 9.2   & 36.9  & 29.4   \\ %352.9+11.4
IC1297            &   3   & 0.4   &  2.9  &  5.7   \\ %358.3-21.6
\hline
\end{tabular}
\end{center}
\end{table}

Figure 1 shows the IRAS colour--colour diagram for the objects of
Table 1.  In Fig. 1a the 40 [WC] stars (filled circles) are shown
together with all other PNe (open circles). The full PNe sample was
selected as all PNe in the ESO-Strasbourg catalogue of PNe, with flux
quality of 2 or 3 in all three bands, and where the IRAS position is
within 20 arcsec of the optical position in the catalogue.  The [WC]
and other object cover the same region of the diagram, but the [WC]
stars are more predominant among the bluer objects.  The [WC] stars
have average IRAS colours of (0.75,0.01) in Fig. 1a, while the
non-[WC] PNe have on average (0.85,0.02). Thus, the [WC] stars tend to
have a relatively stronger $F_{12}$ flux ratio but the $F_{60}/F_{25}$
is no different.  Sczcerba (these proceedings) shows that ISO spectra
of some [WC] objects show faint PAH features: these could raise the
12-micron flux and cause a blueward shift in the diagram. It thus appears
that the dust characteristics differ between [WC] and non-[WC] stars.

Figure 1b shows the IRAS colours as function of [WC] subclass, where
the size of the circles is proportional to the [WC] subclass; i.e.
the largest circles correspond to the coolest central stars.  The
coolest stars have the hottest dust, implying their dust is nearest to
the star. This agrees with a general evolution  towards hotter stars
while the nebula expands: the [WC] stars evolve towards higher
temperatures on time scale comparable to the nebular expansion. The
difference in distribution in Fig. 1a appears to be caused by the
cooler [WC] stars only.

The boxes in Fig 1b indicate the typical location of different
classes of objects, as indicated in the Figure caption.  The [WC]
stars are found on the blue side of the PNe box (as discussed before),
with several having colours typical for young Post-AGB stars, with
nebular time scales of only a few hundred years. A few [WC] stars have
a 60-$\mu$m excess which puts them in a region where objects with
bipolar outflows are found (Zijlstra et al. 2001).

In Fig. 1c the size of the open circles is proportional to the log of
the integrated IRAS flux.  The integrated flux is obtained by summing
the in-band flux of the three bands and is thus a
lower limit to the integrated dust emission, especially for hot
objects.  The objects with the hottest dust are also
very bright in IRAS. This trend is affected by the unknown distances:
the flux will scale with $d^{-2}$ where $d$ ranges from perhaps 1 kpc
for the nearest objects to 8 kpc for Bulge nebulae. However, the dust
emission declines so strongly as the nebula expands that the distance
spread is a secondary parameter. 

The dots in Fig. 1c are all entries in the IRAS point source catalogue
with flux quality 3 at 12, 25 and 60$\mu$m, and with total in-band
flux $F_{ir}>10^{-11}\,\rm W\, m^{-2}$ which is similar to the
brightest [WC] stars. There are very few such IRAS sources especially
in the PNe region, and those show colours clearly distinct from the [WC]
stars. The bright [WC] stars form a well defined group among IRAS
sources with little confusion with other types of objects. For
comparsion, Fig. 1d shows all IRAS sources with flux quality 3 at 12,
25 and 60$\mu$m (for all values of total flux). The AGB and its
termination point are clearly visible, with the bright [WC]
stars located near this point.

There is a gap in the distribution of bright IRAS sources in Fig 1c,
at ($0.5,-0.3$). This gap corresponds to the AGB--Post-AGB transition
where evolution is extremely fast ($\sim 100\,\rm yr$), when the
envelope first becomes detached. The bright [WC] stars are seen to
straddle this gap.

\section{Defining the IR-[WC] stars}

\begin{figure}
\includegraphics[width=80mm,height=70mm]{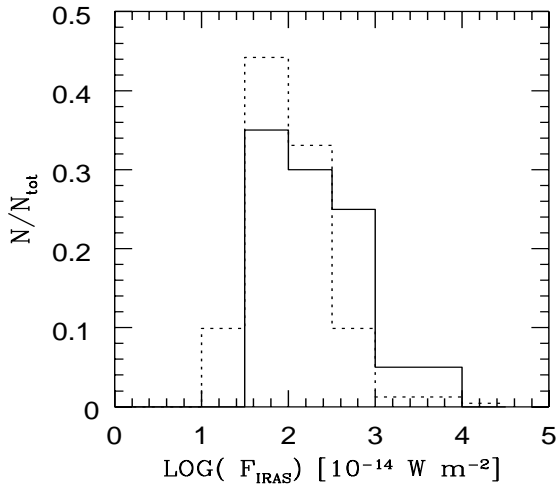}
%\vspace{7cm}  % amount of vertical space needed
\caption{Histogram of the distribution of total in-band IRAS flux,
for [WC] stars (drawn line) and other PNe (dashed line), for
all objects with IRAS detections at 12, 25 and 60$\mu$m}
\end{figure}

Figure 2 shows the distribution of total in-band IRAS fluxes, with the
drawn line representing the [WC] stars and the dashed line all
non-[WC] PNe of Fig. 1a.  The two distributions peak at a similar flux
($10^{-12}\rm \, W\, m^{-2}$), but the [WC] stars dominate at the
highest integrated fluxes.  The brightest IRAS PNe are almost all [WC]
stars. These [WC] stars are the objects which are located near the
post-AGB gap of Fig. 1c; all have late subtypes, or cool central
stars. These are the IR-[WC] stars, defined as [WC] stars with
IRAS in-band flux $ F_{ir} > 0.8 \times 10^{-11} \,\rm W\,m^{-2}$ and
IRAS colours near or in the post-AGB gap.

In Table 1, the IR-[WC] stars are indicated either as confirmed or
possible.  Confirmed stars (BD+30$^{\rm o}$3639, IRAS07027$-$7934, He
2-113 and He 3-1333 (also known as CPD$-$56$^{\rm o}$8032) ) have
colours near or in the gap and IRAS in-band flux $ F_{ir} > 0.8 \times
10^{-11} \,\rm W\,m^{-2}$. The suspected candidates have similar
colours but lower flux (possibly because of a larger distance).  The
[WC] subclasses range from 9 to 11, with one possible candidate
(M2-43) having 8. This very limited range of subclass is not part of
the selection criterium but indicates the likely common origin of the
IR-[WC] stars.

\begin{figure}
\includegraphics[width=125mm,height=70mm]{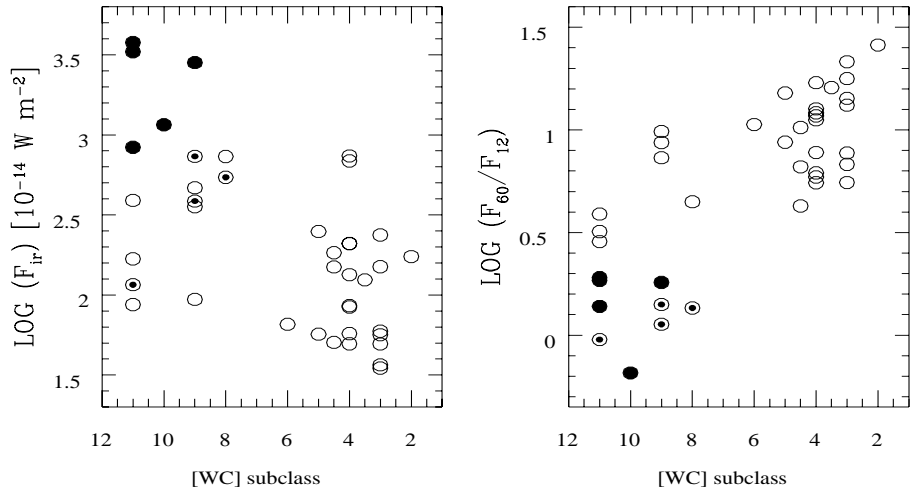}
%\vspace{7cm}  % amount of vertical space needed
\caption{Total in-band IRAS flux and 60/12 micron colour versus [WC]
subclass for the IR-[WC] stars (filled circles), possible IR-[WC] stars
(half-filled circles) and other [WC] stars (open circles) }
\end{figure}

Fig. 3 shows the IR-[WC] stars (filled circles), the possible ones
(partly filled circles) and the other [WC] stars of Table 1 (open
circles).  The IR-[WC] stars very clearly separated from the other
[WC] stars.  The gap at intermediate subclasses (5--7) is clear (Gorny
\& Stasinska 1995).  The IR-[WC] stars all belong to the late [WC]
stars. However, not all late [WC] stars belong to this class.  Leaving
out the IR-[WC] stars, the gap at subclasses 5--7 is much less
significant.

The right panel effectively measures the distance in the
colour--colour diagram from the AGB, showing the IR-[WC] stars should
be in an extremely early post-AGB phase. These colours are however very
different from those of the well-known bipolar post-AGB stars (such
as, the Egg nebula and similar objects; Zijlstra et al. 2001) which
are found near region V of Fig. 1b.

{\it These IR-[WC] stars with colours similar to young post-AGB stars
have few or no counterparts among the non-[WC] PNe.} They define a
separate class of objects, and point at a genuine difference in
nebular evolution between some late [WC] stars and other PNe.

\section{The silicate problem}

Zijlstra et al. (1991) found an OH maser originating from the [WC11]
star IRAS07027$-$7934. The IRAS LRS spectrum shows strong PAH
features. This indicated a chemical dichotomy with both carbon-rich
and oxygen-rich regions being present. The OH maser showed strong
1612MHz emission, a transition which requires long path lengths at
moderate temperatures (Field 1985). The PAH emission require a strong
UV field. This implies that the oxygen-rich material is situated in
the outer regions, while the carbon-rich dust is found close to the
star. This was seen as evidence for a very recent (less than 500 yr
ago) change of the star from oxygen to carbon rich, probably related
to the thermal pulse. Thus, IRAS07027$-$7934 appears a candidate for a
[WC] star which can only have formed through a fatal thermal pulse
scenario.

\begin{figure}
\includegraphics[width=125mm]{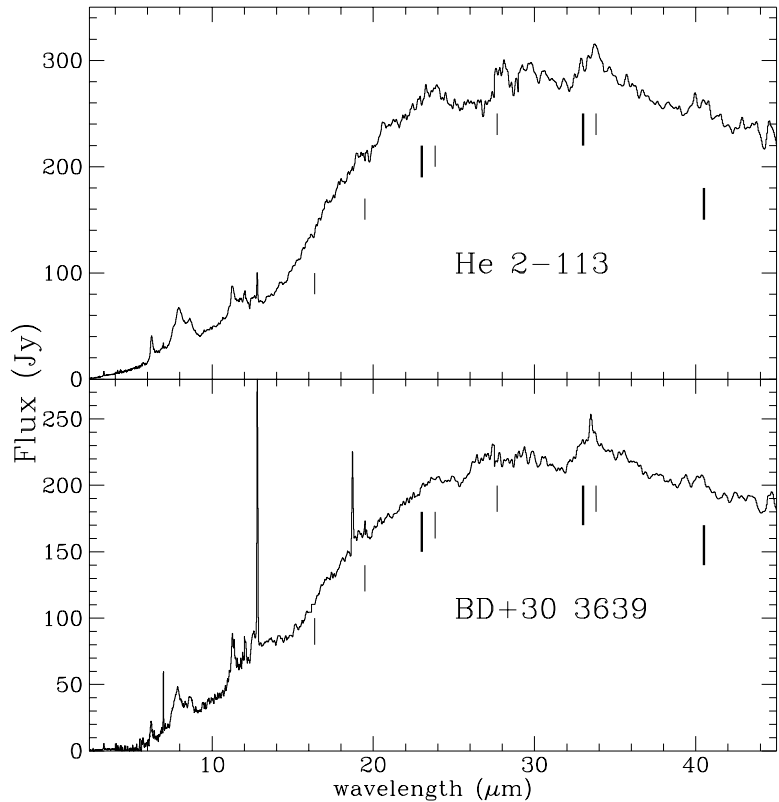}
%\vspace{14cm}  % amount of vertical space needed
\caption{ISO spectra of two IR-[WC] stars, showing the PAH features
at $\lambda < 13\mu$m, and the crystalline silicate bands at   
$\lambda > 20\mu$m. Thin marks indicate olivine bands and thick marks
indicate pyroxenes (from Waters et al. 1998). }
\end{figure}

Although previously a unique case, ISO spectroscopy found a similar
chemical dichotomy in other [WC] stars, with strong PAHs bands at
short wavelengths and bands of crystalline silicates (olivines and
pyroxenes) at longer wavelength. This has been seen in He 2-113 and
BD+30${\rm ^{o}}$3639 (Waters et al. 1998) and in He 3-1333 (Cohen et
al. 1998). The temperature of the crystalline dust is lower than the
PAH particles, indicating there is a carbon-rich region close to the
star but an oxygen-rich region at larger distances. In BD+30${\rm
^{o}}$3639 and He 3-1333 the ionized region is known to be strongly
carbon rich and the likely seat of the PAH emission (Siebenmorgen et
al. 1994).  The chemical change in the stellar wind was dated to about
1000 yr ago for both He 2-113 and BD+30${\rm ^{o}}$3639, and 250 yr
for He 3-1333, consistent with the estimates for IRS07027$-$7934.
In all cases, the silicates appear to be located in the
molecular part of the envelope.\footnote{For some relevant results after these
  proceedings were published (2001), see Toala et al. 2019, Guzman-Ramirez et al. 2015,
  Clayton et al. 2011, Garcia-Hernandez et al. 2008} 

The observations show that probably all confirmed IR-[WC] stars show
the double chemistry. IRAS 21282+5050 has not yet been observed at
long wavelengths, but it also shows very strong PAH emission (Geballe
et al. 1994).  

Chemical dichotomy (showing both a carbon-rich and an oxygen-rich
region) is very rare in post-main-sequence objects is. Novae show this
behaviour (Evans et al. 1997). The Red Rectangle (Waters et al. 1996)
shows PAHs band and strong crystalline silicates.  IRAS08005$-$2356
(Bakker et al. 1997, Zijlstra et al. 2001) shows carbon molecules and
an OH maser. Both the last two stars are too cool to ionize their
nebulae. They are binary stars with the crystalline silicates trapped
in an ancient disk: they are probably not related in any way to the
[WC] stars.  Roberts 22 also has PAHs, an OH maser and crystalline
silicates. It has an inner ionized region (Zijlstra et al. 2001) but
is probably a supergiant (Molster, priv comm.). Among PNe, NGC 6302
has very weak PAH emission and crystalline features.

It appears likely that the presence in PNe  of strong PAH features in
combination with OH masers and/or crystalline silicates is
uniquely related to the IR-[WC] stars. \footnote{Work done since these
  proceedings have shown that weak PAH emission is
  common in planetary nebulae, both oxygen-rich and carbon-rich, but
  strong PAHs are rare and indicate carbon-rich nebulae: Cerrigone et
  al. 2009, Guzman-Ramirez et al. 2014. Although the IR-[WC] stars are
strong PAH emitters, they do not appear to show fullerene emission:
Otsuka et al. 2014}

% The reason why a preculiar nebula is associated with late [WC] stars
% is not clear, especially since the PAH components appear to have
% formed {\it before} the star became hydrogen poor.  The nebular PAH
% emisison is more than a reflected history of the star, but points at a
% on-going connection between star and nebula.

\section{Time scales}

The IR colours, the IR fluxes and the chemical dichotomy all point at
very short time scales of the order of $10^2$--$10^3\,$yr.  The
apparent conclusion is that the IR-[WC] stars were on the AGB before
that time. The time when the star became hydrogen-poor may be even
more recent: BD+30${\rm ^{o}}$3639 shows evidence for hydrogenated
PAHs (Waters et al. 1998) implying that the PAHs formed after the star
became carbon rich but before it became hydrogen poor. 

Cohen et al. (1998) note that such short time scales are implausible
from evolutionary considerations, and suggest that the silicate grains
do not participate in the nebular expansion and date back to a much
earlier epoch.  Support for their argument comes from a separate
reasoning: if the timescale are so short, the stars should evolve
rapidly in the HR diagram. One would expect to see hotter stars with a
similar chemical dichotomy. Such stars are not known.  On the other
hand, hydrogen-poor stars must have very low stellar envelope mass, and the cool
[WC] stars have wind with very high mass-loss rates.  (Note that the
extreme C/O ratio of He 3-1333 and He 2-113 (de Marco et al. 1998)
suggest a very low envelope mass (e.g. Frost et al. 1998)).  Low
envelope mass would lead to very fast evolution towards higher
temperatures. The lack of hot IR-[WC] stars is therefore puzzling.

The number of IR-[WC] stars can be used to obtain limits on the
evolutionary age of the IR-[WC] stars. The IRAS in-band flux is about
$10^{-11}\rm \,W\, m^{-2}$ which corresponds to a luminosity of
approximately $10^{29} d^2 \rm\, W$ where $d$ is the distance in kpc.
The in-band flux will underestimate the total IR flux typically by a
factor of 2 for objects with hot dust. This gives $L_{\rm IR} \approx
500\, d^2 \,\rm L_\odot$. The distance should be less than about 3 kpc
for post-AGB luminosities.  In fact distance estimates for IR-[WC]
stars are mostly 1--2 kpc.

The local column density of PNe is about 25 kpc$^{-2}$ (Zijlstra \&\
Pottasch 1991).  The IR-[WC] PNe account for 1--2\%\ of all PNe.  For
a life expectancy of $3\times 10^4\,\rm yr$ for a PN (Zijlstra \&\
Pottasch 1991), we can expect one object within 2 kpc for every 100-yr
phase of PN evolution.  The number of IR-[WC] stars than implies a
minimum life time of 500--1000 yr for a IR-[WC] PN if {\it all} PNe
pass through this phase.  The oberved fraction of [WC] stars (8\%) can
be used as an upper limit to the fraction of PNe which may have been
IR-[WC] stars. We find a {\it minimum} life time for an IR-[WC] star
of $5\times 10^3\,\rm yr$, with a likely life time in excess of
$10^4\,\rm yr$.

This is in contradiction with the implied short nebular time scales,
and agrees with the suggestion of Cohen et al. The discrepancy can be
reduced by assuming a higher distance and luminosity for the IR-[WC]
stars. However, PNe with stellar luminosity $> 10^4 \,\rm L_\odot$ are
very rare and would evolve rapidly, again decreasing the expected
numbers.  The observed number of objects shows that IR-[WC] stars are
comparatively long-lived.  {\it Both nebulae and stars must evolve
much slower during the IR-[WC] phase than their compact nebulae suggest.}

\section{Morphology}

A way to slow down evolutionary time scales of the nebulae is by
assuming the existence of a slowly expanding torus or disk (e.g. Waters
et al 1998). This would lead to bipolar morphologies, with
possibly high velocities perpendicular to the torus.

BD+30$\rm ^o$3639 is among the most spherically
symmetric PNe known. However, Bryce \&\ Mellema (1999) have shown that
in fact it has a high-velocity bipolar outflow and thus is likely
bipolar. The nebula is seen almost pole-on hiding its true structure.

CPD$-$56$\rm ^o$8032 and He2-113 are very compact but HST images have been
published by de Marco et al. (1997). Both show irregular but 
ring-like structures. The structures are consistent with bipolar
geometries but do not proof it. Curiously, both objects show that
the 'central' star is in fact located off-centre of the nebulosity.
The ionized core of IRAS 07027$-$7934 has not been imaged, being
subarcsecond.  There are a few indications of asphericity (the split
OH profile, the fact that the CO expansion velocity is half that of
the OH, and the fact that the ionized core is not quite at the centre
of the H$\alpha$ scattering halo, Zijlstra et al. 1991), but no conclusive
evidence. IRAS 21282+5050 has a strong bipolar morphology (Bregman
et al. 1992). \footnote{More recent morphological work is in
  Garcia-Hernandez et al 2006}

The evidence is at present inconclusive, but bipolar, torus-like
morphologies, are possible for the IR-[WC] nebulae; these would allow
for slower nebular expansion.  Note that similar bipolar morphologies
are seen in 10--15\%\ of PN (Manchado et al. 1996). Off-centre central
stars are however very rare.

\section{Extra galactic [WC] stars: the effect of metallicity}

There are few or no Bulge [WC] stars in Table 1, as argued above.
However, there are [WC] stars known in nearby galaxies of which three have
IRAS detections: He2-436 in the Sagittarius dwarf spheroidal at 25
kpc (Sgr: Zijlstra \&\ Walsh 1996), and SMP 58 and SNP61 in the LMC
(Zijlstra et al. 1994). Their IRAS data is listed in Table 2 where the
flux densities for He 2-436 were re-determined using IRAS Software
Telescope data (SRON, Groningen).

\begin{table}[htb]
\begin{center}
\caption{IRAS data for extra-galactic [WC] stars}
\begin{tabular}{llllll}
\hline
Name & [WC] subtype & $F_{12}$ & $F_{25}$ &  $F_{60}$  \\
\hline
He 2-436 (in Sgr) & 3/4 & 0.22 Jy & 0.49 Jy & 0.11 Jy \\
SMP58 (in LMC) &   4/5 & 0.15 Jy & 0.22 Jy & -- \\
SMP61 (in LMC) & 4/5 &    0.08 Jy & 0.13 Jy & $<$0.16 Jy \\
\hline
\end{tabular}
\end{center}
\end{table}

Sgr has a metallicity of [Fe/H]$=-0.55$ and a main-sequence turn-off
mass of 1.3 M$_\odot$ (Dudziak et al. 2000). Its two [WC] stars are
the only such stars where these parameters are known.

The IR colours and IR flux (when scaled to a distance of 2 kpc) are
similar to those of the IR-[WC] stars. However, Fig.  5 shows their
[WC] subclass is much earlier than for Galactic stars with similar IR
characteristics. Possibly the stars evolve faster to higher
temperature at lower metallicity. Another possibility is that the
subclass is affected by metal blanketing in an extended wind, in which
case lower metallicity causes the $\tau=1$ surface to be closer to the
real stellar surface. (This would however imply that the late IR-[WC]
stars  have much hotter stars than the wind emission lines
suggest.)

\begin{figure}
\includegraphics[width=80mm]{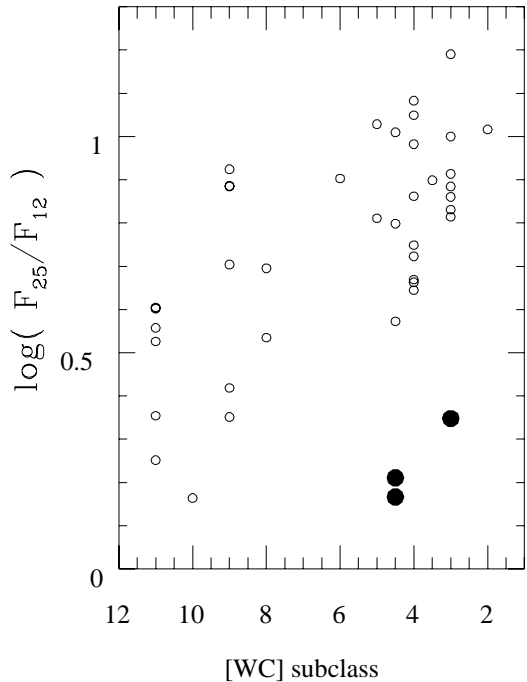}
%\vspace{7cm}  % amount of vertical space needed
\caption{ $F_{12}/F_{25}$ as function of [WC] subclass, for Galactic [WC]
stars (open circles), and the extra-galactic [WC] stars (filled circles) }
\end{figure}

The data suggests that the group of the IR-[WC] stars is not confined
to the Galactic disk. However, the detailed properties show
a dependence on metallicity. Given the AGB populations of Sgr and the LMC, it
is likely that their progenitor stars were already carbon-rich before
the ejection of the PN on the AGB. A chemical dichotomy similar to
the Galactic IR-[WC] stars is therefore not expected.

\section{The problem of evolution}

There are three main problems regarding the IR-[WC] stars: (1) A
unique nebular double chemistry seen {\it only} around IR-[WC] stars;
(2) The conflict between the very short time scales implied by the
nebulae and by the low envelope mass expected for a hydrogen-poor
Post-AGB star, and the much longer timescales implied by other
considerations; (3) The fact that the dense nebulae favour a
fatal-thermal pulse (FTP) scenario whereas stellar model calcuations
favour a VLTP to obtain hydrogen-poor stars.

Long time scales require slow post-AGB evolution of the star.  Slow
evolution is expected for stars with very low core mass. However, their
luminosity would also be very low, which is not observed. 

The only other way to ensure slower post-AGB evolution is
by replenishing the envelope of the star, to compensate for the
effects of nuclear burning and wind mass loss. This would
require accretion of mass from the ejecta back onto the star.

An accretion scenario could explain the link between dense, compact
nebulae and the peculiar evolution. It requies that the hydrogen
is efficiently mixed down in the star to maintain the [WC]
characteristics. In such a model, the PAHs molecules could in fact
form in the accreting gas, where CO is dissociated (as in novae,
Evans et al. 1997). Furthermore, slow accretion could trigger
a VLTP on the cooling track where otherwise this might not occur.

Such a model is highly speculative. It is however clear that the
FTP and VLTP scenarios, while adequate for the other [WC] stars,
do not as yet provide a good explanation for the IR-[WC] stars.

\end{document}